\documentclass[fleqn,usenatbib]{mnras}

\usepackage{newtxtext,newtxmath}

\usepackage[T1]{fontenc}
\usepackage{ae,aecompl}

\usepackage{graphicx}	
\usepackage{amsmath}	
\usepackage{amssymb}	

\newcommand{\mrm}[1]{\ensuremath{\mathrm{#1}}}

\newcommand{\GX}{GX~339$-$4}
\newcommand{\oi}[1]{OS#1}

\newcommand{\RXTE}{\textit{RXTE}}
\newcommand{\MJD}[1]{MJD #1}

\newcommand{\ti}[1]{\textit{#1}}

\title[Superhump period of GX~339$-$4]{Superhump period of the black hole X-ray binary GX~339$-$4}

\author[I. A. Kosenkov and A. Veledina]{
Ilia A. Kosenkov$^{1,2}$\thanks{E-mail: ilia.kosenkov@utu.fi, alexandra.veledina@utu.fi} and Alexandra Veledina$^{1,3,4}$\\
$^{1}$Tuorla Observatory, Department of Physics and Astronomy, University of Turku, V{\"{a}}is{\"{a}}l{\"{a}}ntie 20, FI-21500 Piikki{\"o}, Finland \\ 
$^{2}$Department of Astrophysics, St. Petersburg State University, Universitetskiy pr. 28, Peterhof, 198504 St. Petersburg, Russia\\
$^{3}$Nordita, KTH Royal Institute of Technology and Stockholm University, Roslagstullsbacken 23, SE-10691 Stockholm, Sweden\\
$^4$Space Research Institute of the Russian Academy of Sciences, Profsoyuznaya Str. 84/32, Moscow, 117997, Russia\\
}

\date{Accepted 2018 April 28. Received 201 April 26; in original form 2017 December 15}

\pubyear{2018}

\begin{document}
\label{firstpage}
\pagerange{\pageref{firstpage}--\pageref{lastpage}}
\maketitle

\begin{abstract}
We investigate variability of optical and near-infrared light curves of the X-ray binary \GX\ on a timescale of days. 
We use the data in four filters from six intervals corresponding to the soft state and from four intervals corresponding to the quiescent state.
In the soft state, we find prominent oscillations with an average period $P=1.772 \pm0.003$~d, which is offset from the measured orbital period of the system by 0.7~per cent.
We suggest that the measured periodicity originates from the superhumps.
In line with this interpretation we find no periodicity in the quiescent state.
The obtained period excess $\epsilon$ is below typical values found for cataclysmic variables for the same mass ratio of the binary.
We discuss the implications of this finding in the context of the superhump theory.
\end{abstract}




\section{Introduction}\label{sec:Intro}

The spectral and variability properties of accreting black hole X-ray binaries have been studied since early 1960s.
There are about 60 such sources known in our Galaxy, and every year there is, on average, one new discovered.
The vast majority of these systems are transient low-mass X-ray binaries (LMXBs): they undergo an outburst and then return to quiescence again on the time-scale of weeks to months.
The recurrence time for most of the systems is comparable to, or larger than the time-scale of the X-ray astronomy era, thus most of the systems were observed only once.
However, a few persistent and recurrent systems are identified, allowing a comparison of their properties between outbursts.
The black hole binary \GX\ is among these systems.

\begin{figure*}
	\centering 
	\includegraphics[keepaspectratio, width = 0.95\linewidth]{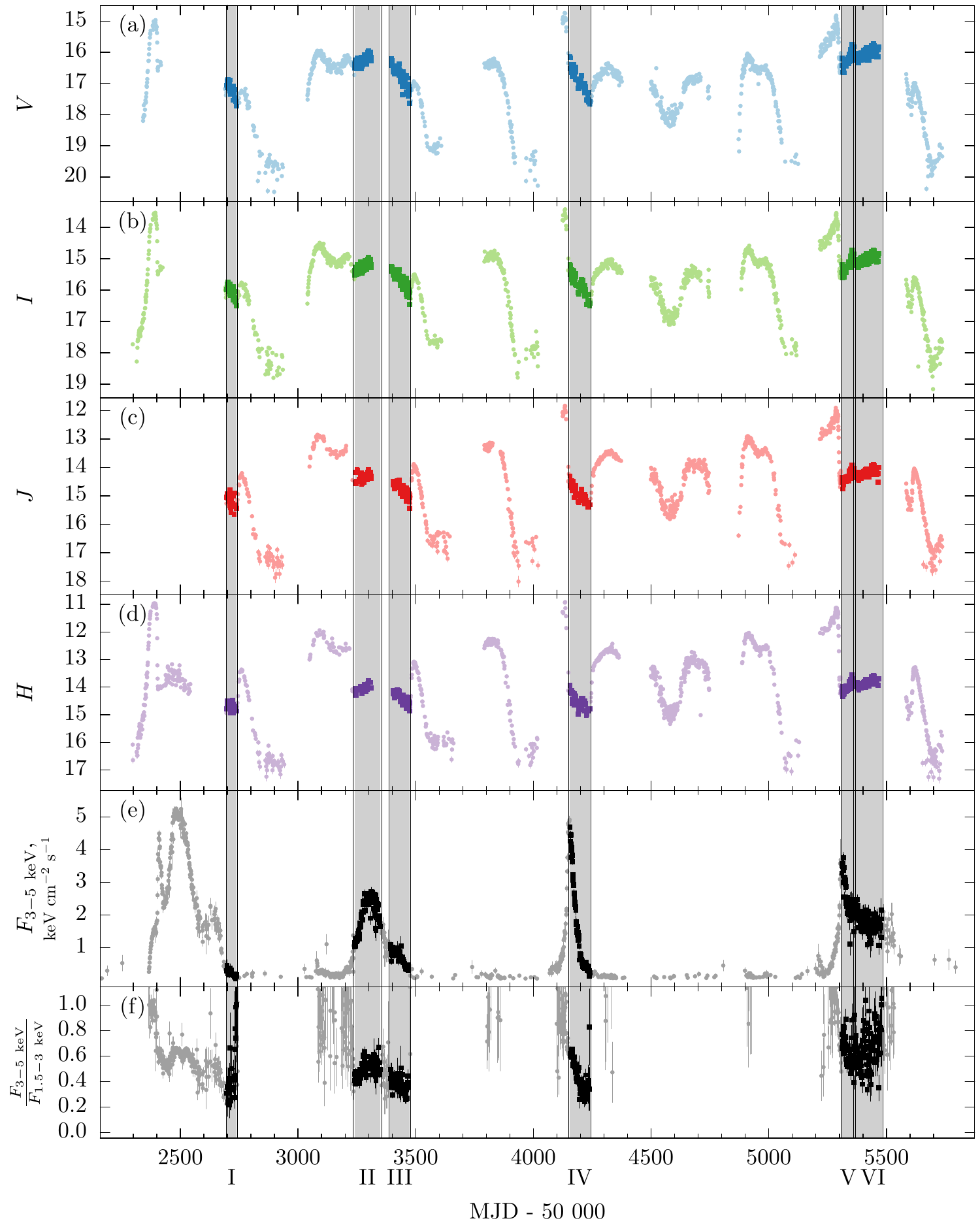}
	\caption{\label{fig:LC}
	(a)-(d) ONIR light curves of \GX\ (the observed magnitudes) as reported in \citet{Buxton2012}. 
	The measurement errors are comparable to the symbol size. 
	(e) \RXTE/ASM B band light curve. 
	(f) ASM hardness ratio (ASM B -- ASM A bands). 
	Grey areas highlight observation sets analysed in this work.
	}
\end{figure*}

The binary undergoes an outburst every 2-3 years and has been observed using multiwavelength campaigns
\citep{SLL99,HBMB05,CB11}.
\GX\ is the standard target for Small and Moderate Aperture Research Telescope System (SMARTS) monitoring and has been observed in the optical and near-infrared (ONIR) using this facility since 2002 \citep[see Fig.~\ref{fig:LC}~a-d;][]{Buxton2012}.
The long and frequent observations revealed the outbursts are proceeding through the sequence of flares, with the flares generally appearing before the transition to the soft state and after the reverse transition \citep{JBOM01,BB04,KDT13}.

Such flares have also been detected in other sources and are believed to arise from the appearance of an additional component, probably of non-thermal origin \citep{CGMR95,JBOM01,BB04,Poutanen2014}.
On the other hand, the ONIR spectra during the soft state seem to approximately agree with the blackbody spectrum, suggesting a major contribution of the irradiated accretion disc.
In the black hole binary XTE~J1550$-$564, the soft-state evolution of the X-ray flux and ONIR magnitudes closely follow the exponential decay profile.
By assuming that the ONIR radiation comes from the X-ray-heated accretion disc, it was possible to obtain the disc temperature using the relationship between the e-folding times in the X-ray and ONIR light curves \citep{Poutanen2014}, providing further grounds for considering ONIR emission as dominated by the disc.
In the soft and intermediate states, the observed magnitudes tightly follow the blackbody track in the colour-magnitude diagram, with only marginal variations.
In contrast, studies of SMARTS \GX\ light curves by \citet{Dincer2012} revealed substantial variability around the mean in the soft state and, partially, during the flare, but not in the quiescent\footnote{The X-ray luminosity during the faintest episodes \citep{ATel2281} is within the luminosity range of formal definition of the quiescent state \citep{McClintock2006}. We refer to these periods as quiescence throughout the paper.} state.
The authors reported the period of soft-state variabilities to be equal to $1.77$~d, which is close to the previously reported orbital period, $P_{\rm orb}\simeq 1.76$~d (see \citealt{Hynes2003}, \citealt{Levine2006}, and \citealt{Heida2017} for a more recent estimate).
Some interesting questions arise: which component in the binary system can produce variability at the orbital period and when can we observe such variations?
Periodic variability can, in principle, come from the moving irradiated surface of the companion, or from the hotspot where the stream of matter from the companion hits the accretion disc, or be caused by the superhumps.
To investigate these questions, we analysed the ONIR data from four outbursts of \GX.

\begin{figure*}
		\includegraphics[keepaspectratio, width = 0.48\linewidth]{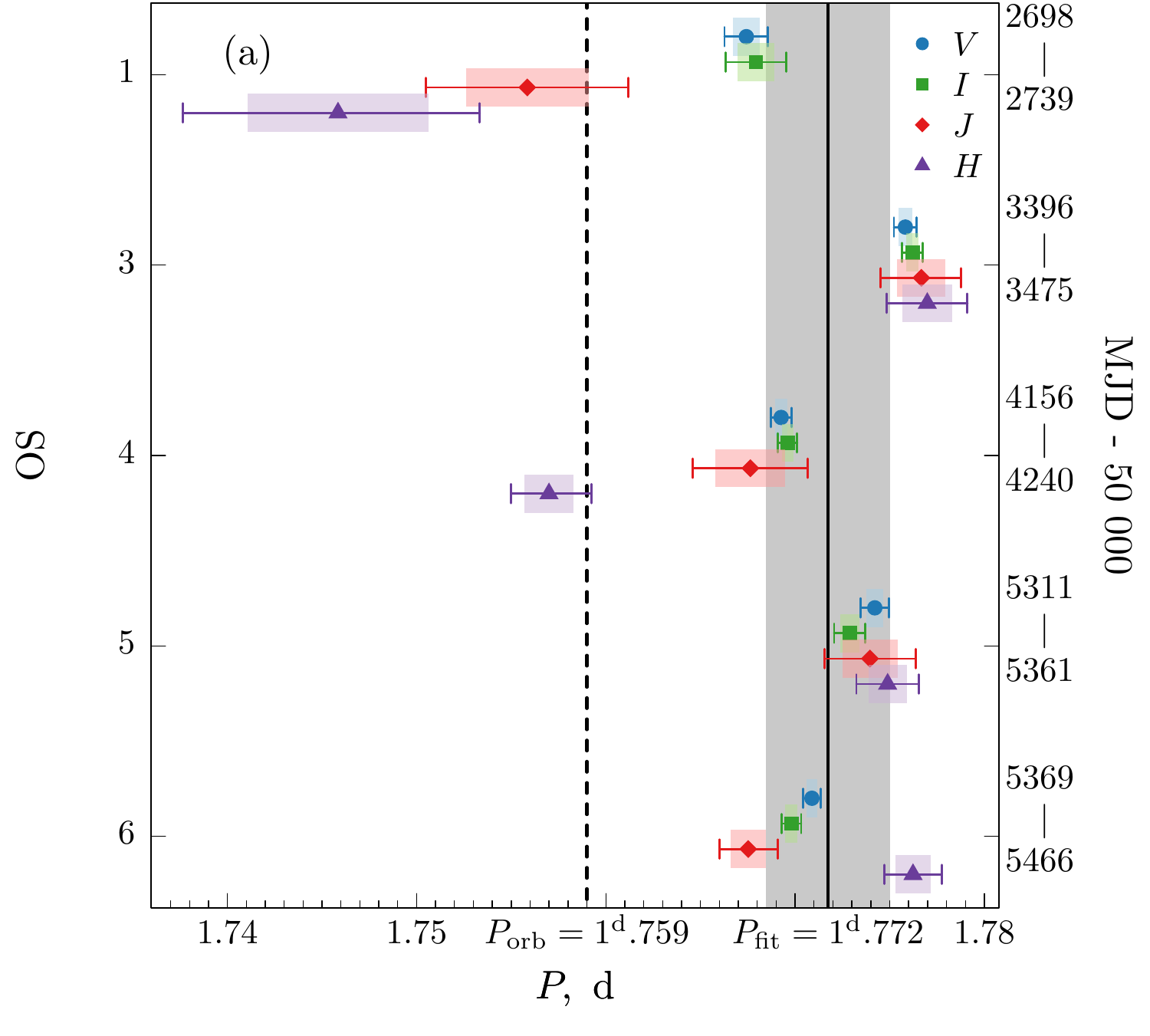}
	\hfill
		\includegraphics[keepaspectratio, width = 0.48\linewidth]{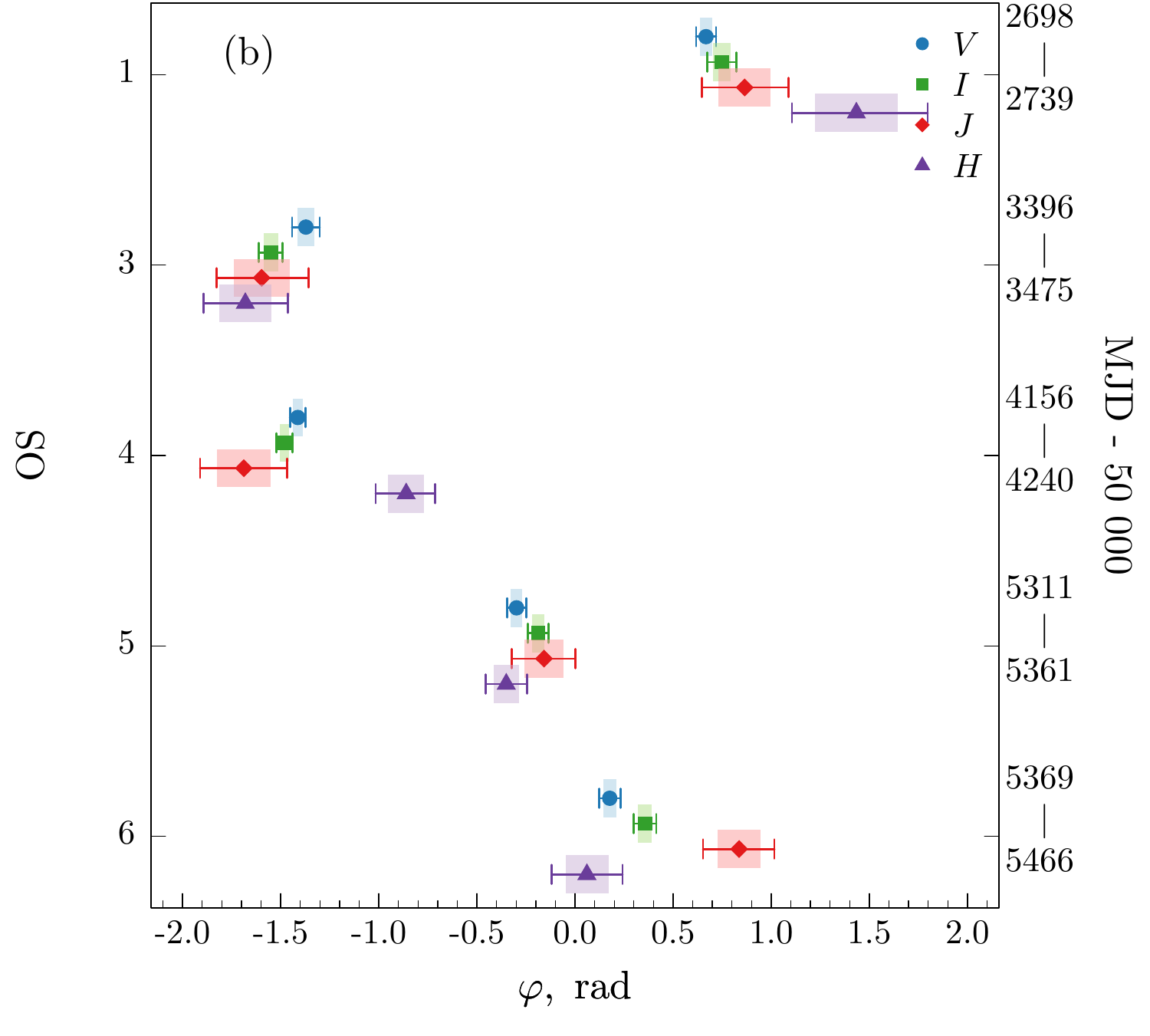}
	\caption{\label{fig:bar_p_ph}
	(a) Distribution of the periods obtained for five intervals using Bayesian inference. 
	The vertical solid line denotes the weighted mean of all shown periods,the grey area indicates the standard deviation of the period distribution.
	The vertical dashed line marks the orbital period of the binary \citep[estimated in][]{Heida2017}. 
	(b) Distribution of the obtained phases. 
	The shaded areas and horizontal bars correspond to $1\sigma$ and $2\sigma$ errors, respectively.
	}
\end{figure*}

\section{Data selection}\label{sec:ONIRDat}
\begin{table}
\caption{\label{tbl:obs_sets}
List of start end dates and number of data points of the observational sets (\oi{s}) analysed in this work.
}
\centering
\begin{tabular}{ccccccc}
\hline
\hline
 \oi      & Start date      & End date   & \multicolumn{4}{c}{N of observations}    \\
			  & MJD$\ -\ 50000$ & MJD$\ -\ 50000$ & $V$ & $I$ & $J$ & $H$ \\
\hline
1 & $2697.85$ & $2738.83$ &  34  &  35  &  31  &  30\\
2 & $3240.46$ & $3310.49$ &  53  &  56  &  24  &  21\\
3 & $3395.86$ & $3474.86$ &  57  &  55  &  56  &  55\\
4 & $4155.88$ & $4239.80$ &  56  &  60  &  47  &  43\\
5 & $5310.62$ & $5360.61$ &  48  &  45  &  42  &  48\\
6 & $5368.64$ & $5466.50$ &  66  &  66  &  60  &  53\\
\hline
\end{tabular}
\end{table}

We use the publicly available SMARTS data\footnote{\url{http://www.astro.yale.edu/buxton/GX339/}} described in \citet{Buxton2012}.
The source was observed in four ONIR bands, \ti{V}, \ti{I}, \ti{J}\ and \ti{H}, between \MJD{52298}\ and \MJD{55836}. 
The observed magnitudes used in the present work are not corrected for extinction.
We select intervals outside of the flares, when we expect the appearance of a non-thermal component.
We exclude the interval \MJD{52400-52550}, because the observations are available in only one band (see Fig. \ref{fig:LC}).
In addition, we separate intervals that show a difference in trend (such as around \MJD{55400}), thus ending up with six observational sets (\oi{1-6}) in each of the four photometric filters (see shaded areas in Fig.~\ref{fig:LC}~a-d).
The start and end MJD dates and the number of data points analysed in each filter of the selected intervals are given in Table \ref{tbl:obs_sets}.

The {\it Rossi X-ray Timing Explorer} All-Sky Monitor\footnote{\url{http://xte.mit.edu/ASM_lc.html}} (ASM) $3-5$~keV light curve and $3-5$~keV/$1.5-3$~keV hardness ratios for the same dates are shown in Fig.~\ref{fig:LC}(e) and (f).
To convert the observed count rate to the energy flux we adopt the procedure based on the assumption of the linear dependence of the energy flux in an ASM band on the count rates from all three ASM bands.
We use the conversion coefficients calculated in \citet{Zdziarski2002}. 
Although the selection was based on the optical data, the resulting intervals predominantly coincide with the source soft states, when we expect the dominant contribution of the accretion disc both in the ONIR and in the X-ray range.

\begin{figure*}
		\includegraphics[keepaspectratio, width = 0.48\linewidth]{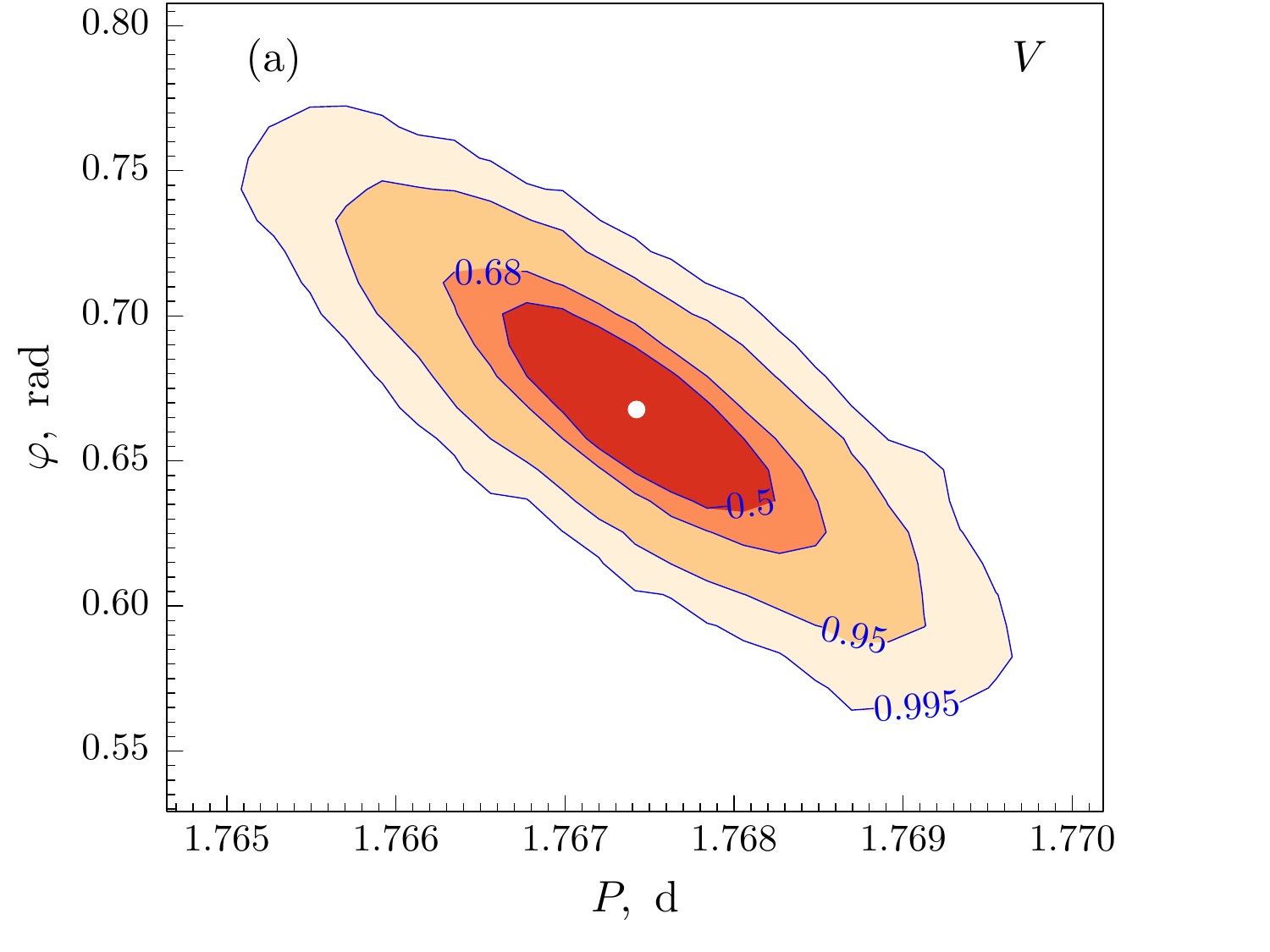}
	\hfill
		\includegraphics[keepaspectratio, width = 0.48\linewidth]{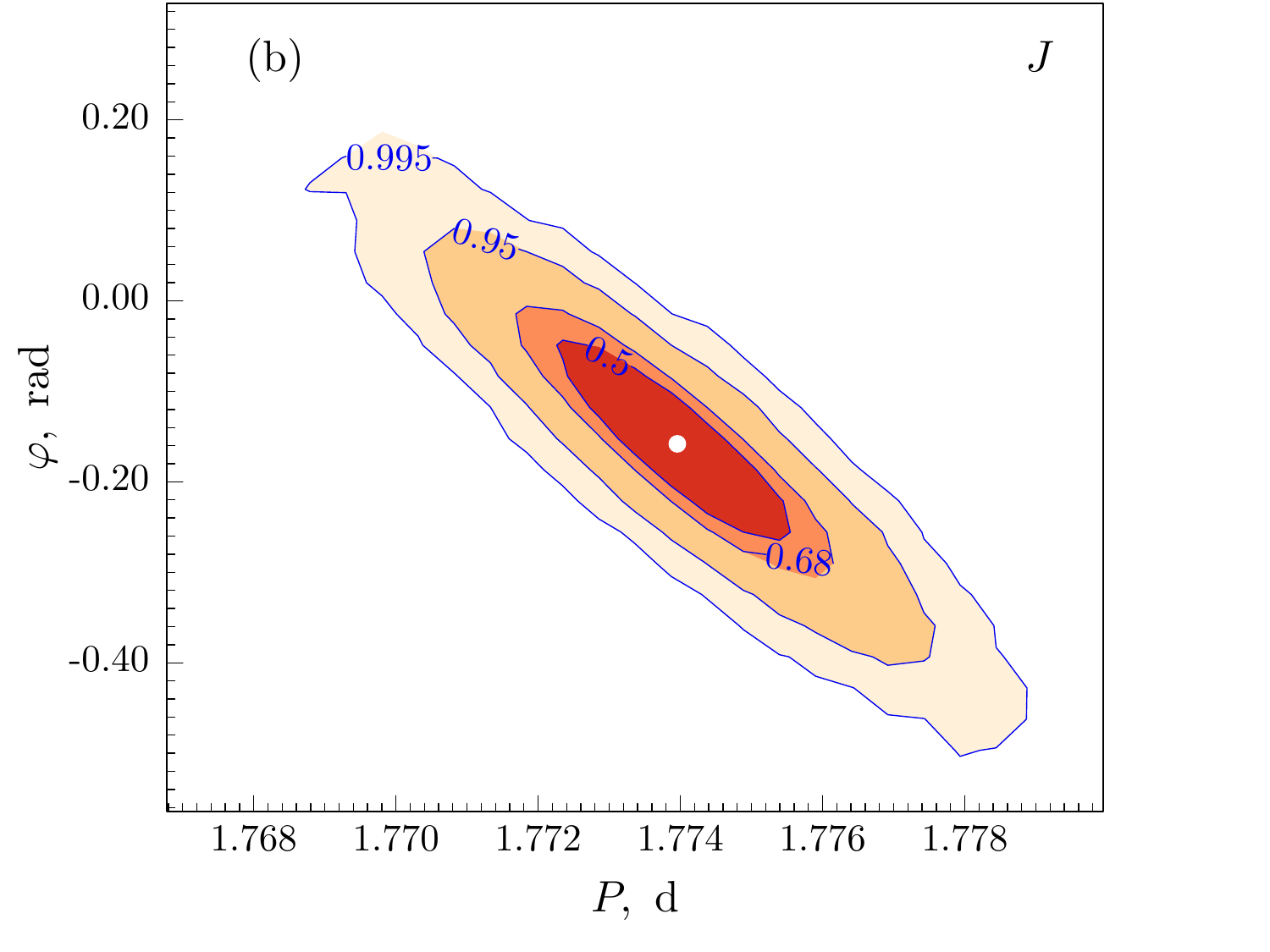}
\caption{\label{fig:dens}
(a) Joint posterior distributions of the period ($P$) and phase ($\varphi$), obtained for the \ti{V}\ band of \oi1. 
(b) Joint posterior distributions obtained for the \ti{J}\ band of \oi5.
Contours represent 0.5, 0.68, 0.95 and 0.995 credibility levels. 
}
\end{figure*}

\section{Data analysis and results} \label{sec:DatAn}
In order to investigate the variability in different intervals and filters, we apply two techniques: the Bayesian analysis of periodicities (Sect.~\ref{sec:DatAn:Bayesian}) and Lomb-Scargle periodograms (Sect.~\ref{sec:DatAn:spectral}).

\subsection{Bayesian inference}\label{sec:DatAn:Bayesian}
ONIR light curves show strong variability on a day time-scale. 
These fluctuations are complemented by the long-term trends that result from the variations on the outburst timescale. 
We assume the ONIR fluxes follow the model
\begin{equation} \label{eq:mdl_flux}
	F_{j}(t) \propto \exp\left(\psi_{j}t\right)\left[1 + \Delta F_{j}\sin\left(\frac{2\pi}{P_{j}}t - \varphi_{j}\right)\right],
\end{equation}
where $j$ corresponds to the various filters (\ti{V}, \ti{I}, \ti{J}\ or \ti{H}), $t$ is the time since the start of the fitted data set (in days), $\psi_{j}$ is the decay rate, $\Delta F_{j}$, $P_{j}$ and $\varphi_{j}$ are the amplitudes, periods and phases of the most dominant modulation, respectively.
The exponential factor is responsible for the trend in the data, and was shown to give a good fit to the similar data on XTE~J1550$-$564 \citep{Poutanen2014}.
The X-ray light-curves are known to be described by the fast rise -- exponential decay profile in the outburst \citep{King1998}, and hence our approximation of the ONIR light curve with exponential decay profile implies an intrinsic connection between these energy ranges (e.g., if the disc is irradiated by the X-ray flux).
We extend this model to the intervals where \GX\ becomes brighter by allowing parameters $\psi_{j}$ to be also positive (\oi2, 5, 6). 

We assume that the variability is caused by the geometrical properties of the source (e.g. varying inclination angle or projected area of the source, or the emission pattern).
In this case, the variability amplitude is proportional to the source brightness; that is, it depends on the brightness multiplicatively.
We show below that the amplitudes $\Delta F_{j}$ take similar values for different outbursts, suggesting common origin of the variability source. 
If the variability is described by an additive model (i.e. the amplitude of variability is independent of flux), we expect variations in quiescence to have higher amplitude than in the soft state, which is not observed. 

In order to simplify the fitting procedure and make the model under discussion suitable for fitting the observed data, we express the relationship in equation~(\ref{eq:mdl_flux}) in terms of magnitudes, keeping the first-order term of the sine component:
\begin{equation} \label{eq:mdl_mag}
  m_{j}(t) = m_{j}^0 + \mu_{j} t  - \Delta m_{j} \sin\left(\frac{2\pi}{P_{j}} t - \varphi_{j}\right),
\end{equation}
where $m_{j}(t)$ is the model magnitude, 
$m_{j}^0$ and $\mu_{j}$ account for the linear trend, and the last term comes from the first-order logarithm series expansion of the modulation component.

We use the Bayesian inference method to estimate the posterior distributions of each parameter. 
We search for periodicities close to the orbital period $P_\mrm{orb} = 1.759$~d \citep{Heida2017} and consider prior periods $P_{j}$ in an interval of $[1.6; 1.9]$~d. 
Other parameters are allowed to vary in a wide range: $\varphi_{j}$ in  [$-\upi$; $\upi$] and $\Delta m_{j}$ in $[0; 0.5]$. 

We process the light curves in each filter of each \oi\ independently, meaning that we obtained 24 sets of parameters after fitting. 
The estimates of model parameters are listed in Table~\ref{tbl:fit_params}.
Systematically smaller observational errors in the \ti{V} and \ti{I} bands lead to smaller error bars for fitted parameters in these filters.\footnote{We note that the observations were carried out simultaneously in two filters, first in the \ti{V}\ and \ti{J}\ pair, then - in the \ti{I}\ and \ti{H} \citep{Buxton2012}, however, we neglect this difference and assume all observations were simultaneous.}

\begin{table*}
\caption{\label{tbl:fit_params}
Values of model parameters estimated using Bayesian inference. 
The parameters $m_{j}^0$ and $\mu_{j}$ determine the linear trend, and $\Delta m_{j}$, $P_{j}$ and $\varphi_{j}$ are amplitudes, periods and phases of the periodic component, respectively. 
The errors correspond to 1$\sigma$. 
} 
\centering
\begin{tabular}{cccccc}
	\hline
	\hline
		 & \multicolumn{5}{c}{Model parameters}      \\
	Band & $m_{j}^0$ & $\mu_{j} \times 10^{-2}$        & $\Delta m_{j}$ & $P_j$  & $\varphi_j$ \\
		 &     & d$^{-1}$ &     & d &  rad       \\
	\hline \noalign{\vskip 1.5ex}
	\multicolumn{6}{c}{OS 1} \\[1ex]
	$V$ &$\phantom{-}16.972 \pm 0.001$ & $\phantom{-}1.334 \pm 0.001$ & $0.135 \pm 0.003$ & $1.767 \pm 0.001$ & $\phantom{-}0.668 \pm 0.031$\\
	$I$ &$\phantom{-}15.829 \pm 0.001$ & $\phantom{-}1.239 \pm 0.001$ & $0.103 \pm 0.003$ & $1.768 \pm 0.001$ & $\phantom{-}0.749 \pm 0.044$\\
	$J$ &$\phantom{-}14.974 \pm 0.001$ & $\phantom{-}0.878 \pm 0.001$ & $0.122 \pm 0.008$ & $1.756 \pm 0.003$ & $\phantom{-}0.865 \pm 0.133$\\
	$H$ &$\phantom{-}14.620 \pm 0.001$ & $\phantom{-}0.614 \pm 0.001$ & $0.090 \pm 0.008$ & $1.746 \pm 0.005$ & $\phantom{-}1.434 \pm 0.212$\\[1.5ex]
	\multicolumn{6}{c}{OS 2} \\[1ex]
	$V$ &$\phantom{-}16.452 \pm 0.001$ & $-0.477 \pm 0.001$ & $0.050 \pm 0.002$ & $1.608 \pm 0.001$ & $-0.299 \pm 0.071$\\
	$I$ &$\phantom{-}15.439 \pm 0.001$ & $-0.461 \pm 0.001$ & $0.065 \pm 0.002$ & $1.744 \pm 0.001$ & $-2.562 \pm 0.048$\\
	$J$ &$\phantom{-}14.288 \pm 0.001$ & $-0.050 \pm 0.001$ & $0.095 \pm 0.008$ & $1.840 \pm 0.001$ & $-3.075 \pm 0.061$\\
	$H$ &$\phantom{-}14.172 \pm 0.001$ & $-0.307 \pm 0.001$ & $0.059 \pm 0.008$ & $1.778 \pm 0.002$ & $\phantom{-}2.985 \pm 0.128$\\[1.5ex]
	\multicolumn{6}{c}{OS 3} \\[1ex]
	$V$ &$\phantom{-}16.322 \pm 0.001$ & $\phantom{-}1.151 \pm 0.001$ & $0.124 \pm 0.002$ & $1.776 \pm 0.001$ & $-1.372 \pm 0.043$\\
	$I$ &$\phantom{-}15.288 \pm 0.001$ & $\phantom{-}0.987 \pm 0.001$ & $0.136 \pm 0.002$ & $1.776 \pm 0.001$ & $-1.550 \pm 0.037$\\
	$J$ &$\phantom{-}14.405 \pm 0.001$ & $\phantom{-}0.838 \pm 0.001$ & $0.123 \pm 0.006$ & $1.777 \pm 0.001$ & $-1.597 \pm 0.142$\\
	$H$ &$\phantom{-}14.081 \pm 0.001$ & $\phantom{-}0.705 \pm 0.001$ & $0.109 \pm 0.005$ & $1.777 \pm 0.001$ & $-1.680 \pm 0.132$\\[1.5ex]
	\multicolumn{6}{c}{OS 4} \\[1ex]
	$V$ &$\phantom{-}16.393 \pm 0.001$ & $\phantom{-}1.344 \pm 0.001$ & $0.129 \pm 0.002$ & $1.769 \pm 0.001$ & $-1.413 \pm 0.024$\\
	$I$ &$\phantom{-}15.400 \pm 0.001$ & $\phantom{-}1.129 \pm 0.001$ & $0.128 \pm 0.002$ & $1.770 \pm 0.001$ & $-1.480 \pm 0.025$\\
	$J$ &$\phantom{-}14.599 \pm 0.001$ & $\phantom{-}0.883 \pm 0.001$ & $0.091 \pm 0.006$ & $1.768 \pm 0.002$ & $-1.688 \pm 0.136$\\
	$H$ &$\phantom{-}14.188 \pm 0.001$ & $\phantom{-}0.910 \pm 0.001$ & $0.127 \pm 0.006$ & $1.757 \pm 0.001$ & $-0.860 \pm 0.091$\\[1.5ex]
	\multicolumn{6}{c}{OS 5} \\[1ex]
	$V$ &$\phantom{-}16.500 \pm 0.001$ & $-1.171 \pm 0.001$ & $0.152 \pm 0.002$ & $1.774 \pm 0.001$ & $-0.298 \pm 0.029$\\
	$I$ &$\phantom{-}15.473 \pm 0.001$ & $-1.157 \pm 0.001$ & $0.135 \pm 0.002$ & $1.773 \pm 0.001$ & $-0.187 \pm 0.032$\\
	$J$ &$\phantom{-}14.554 \pm 0.001$ & $-1.015 \pm 0.001$ & $0.142 \pm 0.006$ & $1.774 \pm 0.001$ & $-0.158 \pm 0.099$\\
	$H$ &$\phantom{-}14.190 \pm 0.001$ & $-0.959 \pm 0.001$ & $0.133 \pm 0.004$ & $1.775 \pm 0.001$ & $-0.350 \pm 0.063$\\[1.5ex]
	\multicolumn{6}{c}{OS 6} \\[1ex]
	$V$ &$\phantom{-}16.163 \pm 0.001$ & $-0.242 \pm 0.001$ & $0.114 \pm 0.002$ & $1.771 \pm 0.001$ & $\phantom{-}0.177 \pm 0.033$\\
	$I$ &$\phantom{-}15.130 \pm 0.001$ & $-0.236 \pm 0.001$ & $0.107 \pm 0.002$ & $1.770 \pm 0.001$ & $\phantom{-}0.356 \pm 0.035$\\
	$J$ &$\phantom{-}14.290 \pm 0.001$ & $-0.243 \pm 0.001$ & $0.092 \pm 0.004$ & $1.768 \pm 0.001$ & $\phantom{-}0.836 \pm 0.110$\\
	$H$ &$\phantom{-}13.949 \pm 0.001$ & $-0.225 \pm 0.001$ & $0.090 \pm 0.004$ & $1.776 \pm 0.001$ & $\phantom{-}0.061 \pm 0.109$\\
	\hline
	\end{tabular}
\end{table*}	

\begin{figure*}
		\includegraphics[keepaspectratio, width = 0.48\linewidth]{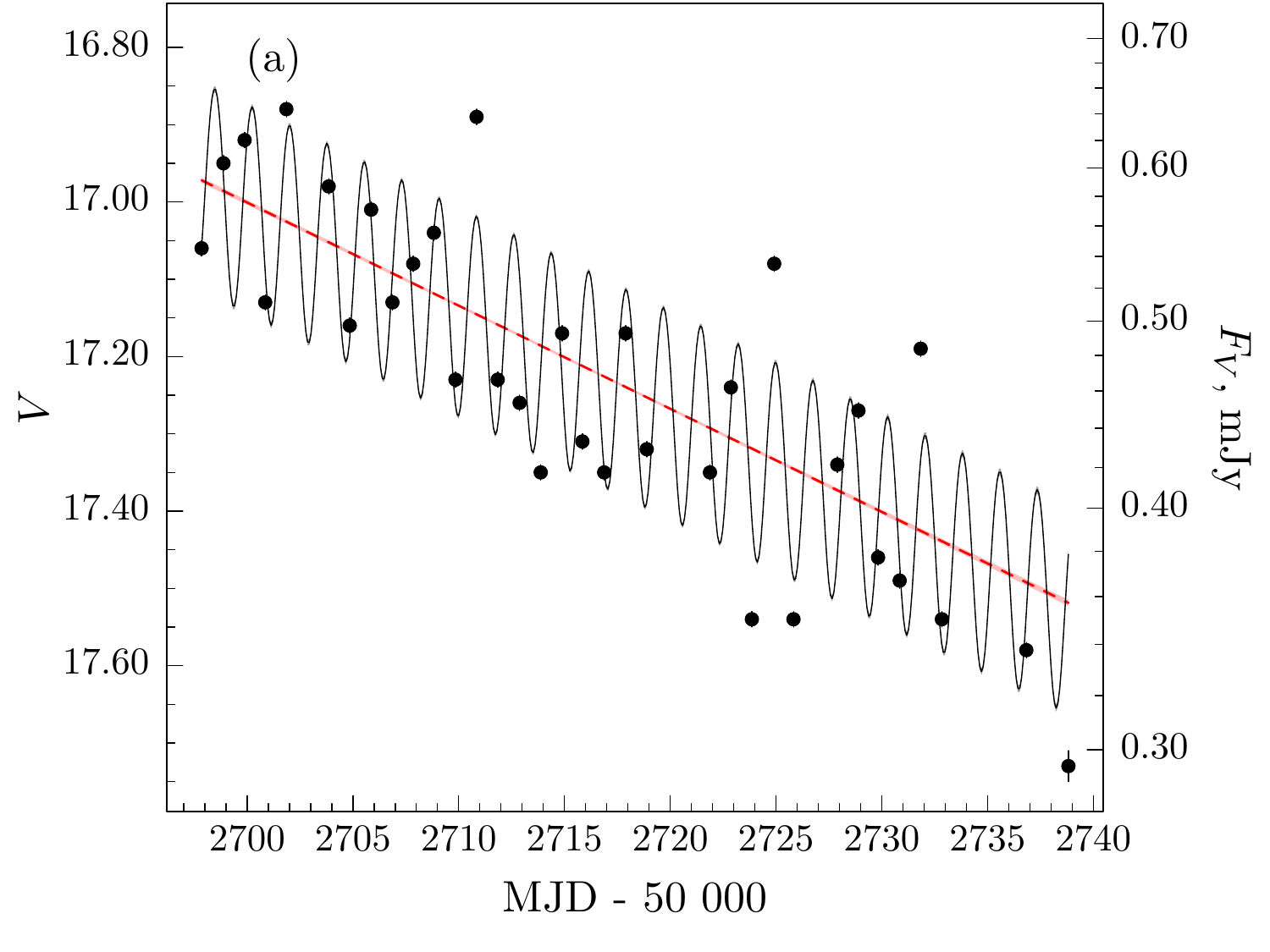}
	\hfill
		\includegraphics[keepaspectratio, width = 0.48\linewidth]{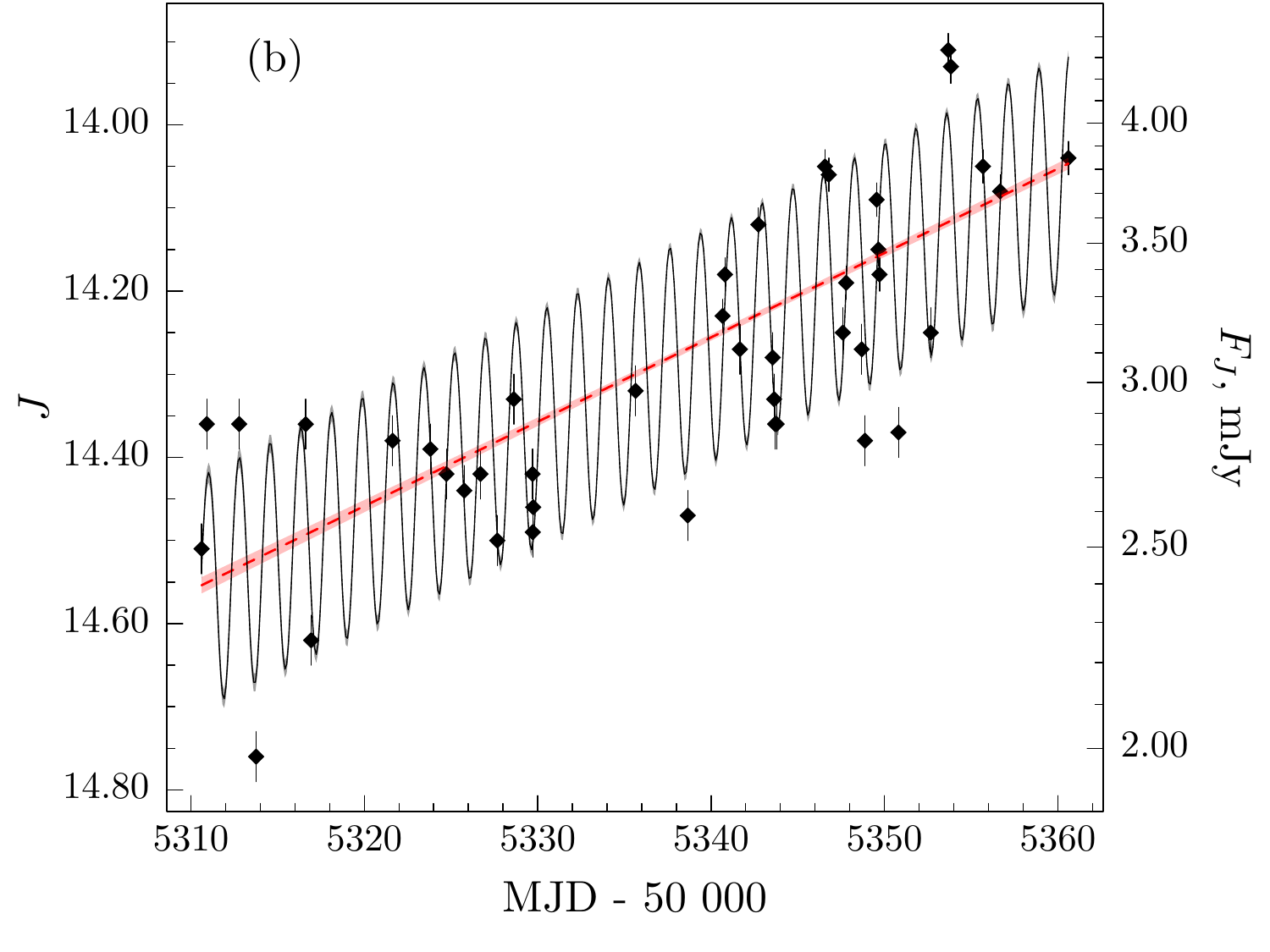}
\caption{\label{fig:fig_models}
The observed light curves in (a) the \ti{V}\ band of \oi1 and (b) the \ti{J}\ band of \oi5.
The trend components are shown with the red dashed lines, and the total model is shown with the black solid line.
Errors are 1$\sigma$; error bars are comparable to the symbol size.
}
\end{figure*}

\begin{figure*}
		\includegraphics[keepaspectratio, width = 0.48\linewidth]{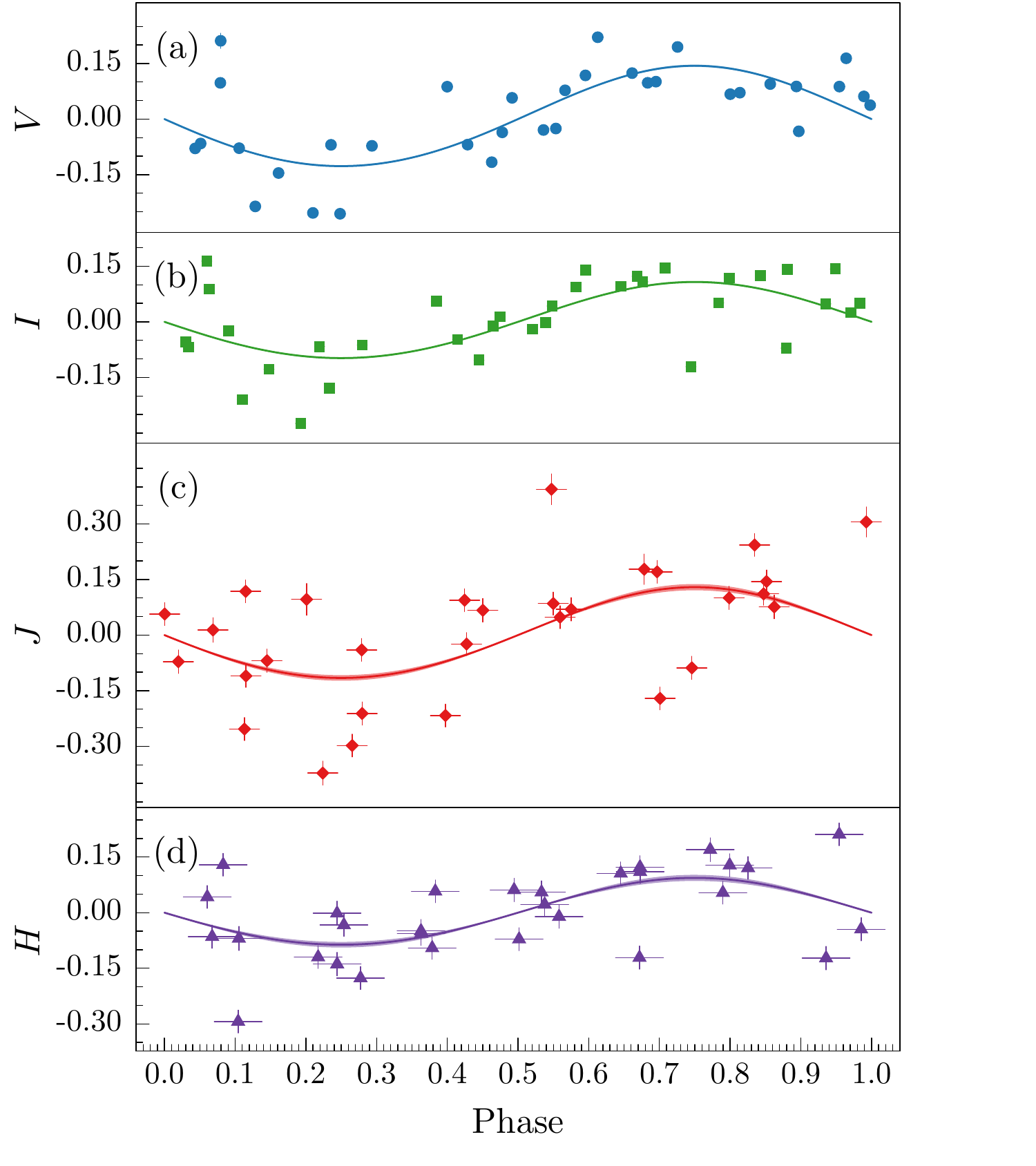}
	\hfill
		\includegraphics[keepaspectratio, width = 0.48\linewidth]{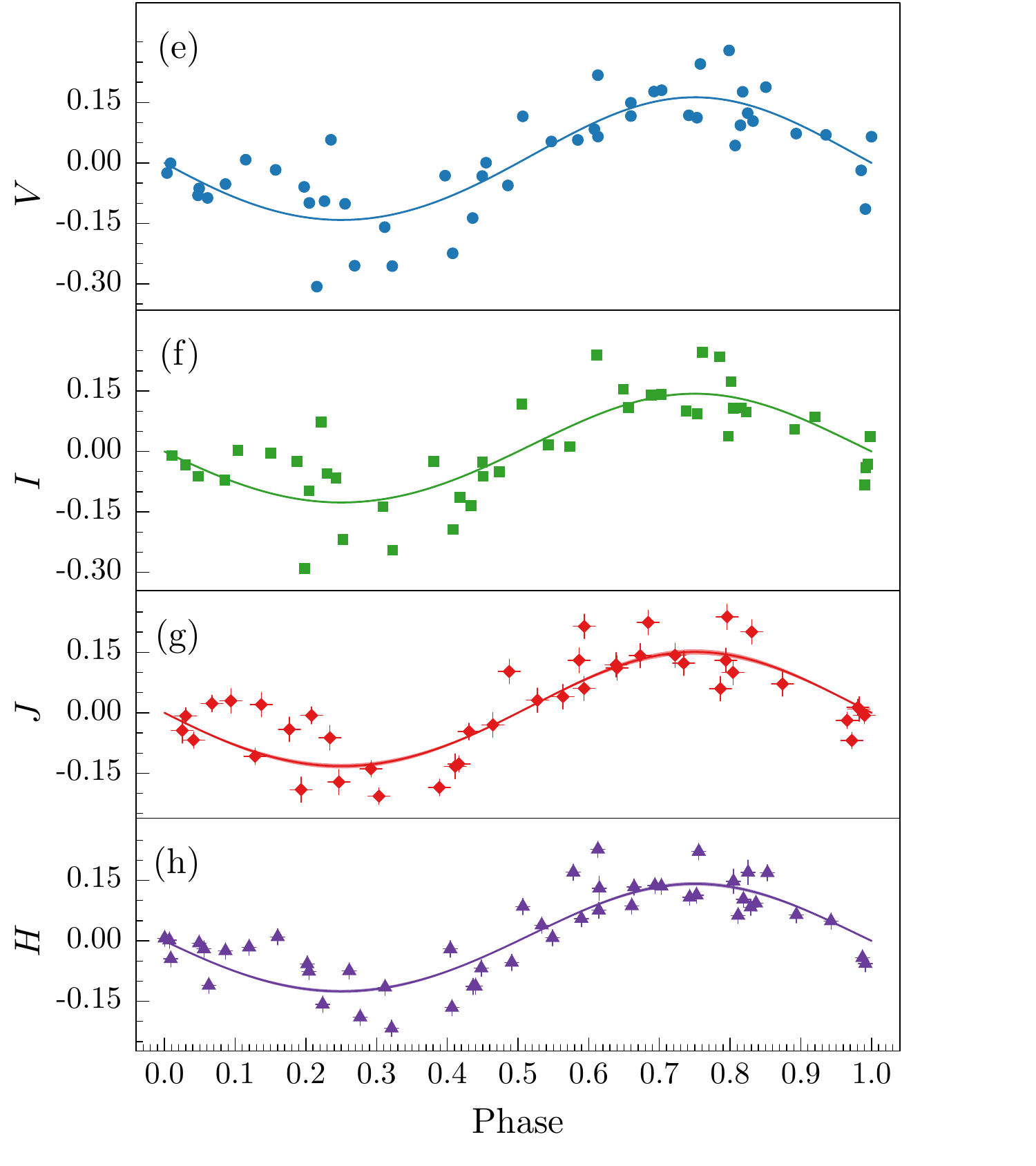}
\caption{\label{fig:folded}
Folded light curves with trend subtracted: (a)-(d) \oi1; (e)-(h) \oi5. 
Horizontal error bars correspond to errors in phase, which are caused by the uncertainties in the estimated model period and phase values.
}	
\end{figure*}

In Fig.~\ref{fig:bar_p_ph} we present the resulting parameters of the periodic component. 
The left panel shows the distribution of the fitted periods and the right panel depicts the respective phases. 
In most cases, the estimates of the model parameters within one \oi\ agree, within 1$\sigma$ errors, in all four filters. 
However, the estimated values of $P$ and $\varphi$ in \oi1 (\ti{J} and \ti{H}) and \oi4 (\ti{H}) are substantially different from the values obtained for other observational sets, because they are influenced by the large errors in observations and weak trends (see Fig \ref{fig:LC}).
In contrast, $P$ and $\varphi$ values of the \ti{V}\ and \ti{I}\ bands tend to be consistent within one observation set. 
We also observe significant evolution of periods from one \oi\ to another, and the difference in periods exceeds the inferred errors. 

\oi2 is absent in Fig.~\ref{fig:bar_p_ph}, because 
the amplitudes obtained for \oi2 are, on average, two times smaller than the amplitudes calculated for other observational sets, and the phases appear to be random (see Table~\ref{tbl:fit_params}). 
This indicates that there is no strong variability in the light curves of \oi2.
Further investigation (see Section~\ref{sec:DatAn:spectral}) of the power spectral density (PSD) of \oi2 light curves supports this result. The PSDs also provide an explanation for the relatively small period errors obtained in \oi2. The Bayesian inference method estimates the parameters of the most prominent modulation, which corresponds to the largest peak in the PSD within the allowed range of prior periods. In the case of \oi2, there are many peaks of similar amplitude (see Fig.~\ref{fig:prdgr_failed}a). Because the peaks are quite narrow, the Bayesian fitting procedure estimates the parameters of these spurious peaks with a reasonable precision. Even though the errors in the period values are small, the amplitudes of the modulations are at the level of error bars, which implies that the obtained periods cannot be trusted.

Examples of joint posterior distributions for periods and phases are shown in Fig.~\ref{fig:dens}(a) (for the \ti{V}-band of \oi1) and Fig.~\ref{fig:dens}(b) (for the \ti{J}-band of \oi5).
The contours are elongated along the direction of $P-\varphi$ anti-correlation.
This is a result of model definition (see equation~\ref{eq:mdl_mag}): setting the argument under the sine function to be constant, we obtain an anti-correlation of $P$ and $\varphi$.

Examples of the observed and modelled light curves are shown in Fig.~\ref{fig:fig_models}(a) (\oi1, \ti{V}-filter) and (b) (\oi5, \ti{J}-filter).
The trend component (red dashed line), assumed to be linear in magnitudes, fits the long-term changes in brightness well. 
The modelled light curve (black solid line) tightly follows the data points, apart from several outliers.
In the \ti{V} band, the deviation of the data from the model is significant (more than $3\sigma$), while in the \ti{J} band the outliers are less than $2\sigma$ away from the model. The nature of the outliers is not clear.

We subtracted the fitted trends from the observed light curves and folded the resulting data points using best-fit estimated periods and phases.
Examples of the folded light curves are presented in Fig.~\ref{fig:folded}(a)-(d) (\oi1) and (e)-(h) (\oi5).
They allow us to investigate the profile of the detected periodic component and deviations from the simple sinusoidal shape. 
There is a hint of the secondary peak in all filters of \oi5 near phase $\varphi = 0.1$, but, owing to the large spread and small number of data points, we cannot draw any firm conclusion. 

The detailed study of profile shape is complicated by the small amount of data points: there are at most 60 data points per 100 days of observations (see Table \ref{tbl:obs_sets}). 
We study the effects of the non-sinusoidal profile on the estimated values of the period by introducing two additional harmonics to the model described in equation~(\ref{eq:mdl_mag}).
We find that the new derived periods are within the errors of the ones estimated with one harmonic, and hence, we see no evidence that the estimated periods are affected by the additional terms in the model. 
We also note that the PSDs obtained using Lomb-Scargle analysis clearly show (in the cases of a confident detection of variability) only one dominant periodic component, and the value of the period at which the PSD reaches maximum agrees with the respective value obtained using Bayesian inference. 

We calculated the weighted mean period for light curves of \oi1, 3, 4, 5, and 6, and obtained $P=1.772$~d (standard deviation 0.003~d).
The weights, chosen to be inversely proportional to the variances of period posterior distributions, allowed us to reduce the contribution of parameters with poor estimates, such as the ones obtained for the \ti{J}\ and \ti{H}\ bands of \oi1. 
However, the weighted average value differs significantly from the orbital periods proposed previously, 
$P_{\rm orb} = 1.7557 \pm 0.0004$~d \citep{Hynes2003, Levine2006} and $P_{\rm orb} = 1.7587 \pm 0.0005$~d \citep{Heida2017}.
This difference cannot be explained by the measurement errors, as the values of both orbital periods and the periods found in our work are obtained with high precision, the latter thanks to the long duration of the observations.
The first spectroscopic orbital period was obtained using data taken during the 2002 outburst \citep{Hynes2003}, about half a year before our \oi1.
More recently, the orbital period was measured during the quiescent state \citep{Heida2017}, almost five years after the last data set analysed in this work was observed.

\subsection{Lomb-Scargle analysis of ONIR light curves}\label{sec:DatAn:spectral}

An alternative method to investigate the periodic components present in the observed light curves is to apply spectral analysis. 
We use the Lomb-Scargle method \citep{Scargle1982} to study the soft-state observations (\oi1 to 6), as well as 
a number of intervals during the quiescent state. 
Unlike the direct fitting of one harmonic, which allows only the most prominent variable component to be detected, the spectral analysis can uncover multiple periodic components of different amplitudes (if present).

In order to apply the Lomb-Scargle method to the observed light curves, we first subtract trends. 
We adopt the trend model similar to the one discussed in Section \ref{sec:DatAn:Bayesian}. 
Preserving the notation introduced in equation~\ref{eq:mdl_mag}, the trend can be approximated as follows:
\begin{equation}
	\hat{m}^\mrm{mdl}_{j}(t) = \hat{m}_{j}^0 + \hat{\mu_{j}} t
\end{equation}
where $\hat{m}^\mrm{mdl}_{j}$ are model magnitudes, and $\hat{m}_{j}^0$ and $\hat{\mu}_{j}$ are linear trend parameters. 
The resulting coefficients can be found in Table~\ref{tbl:spec_params}.
After removal of the linear trend, we applied the spectral analysis procedure to the residuals. 

\begin{table*}
\centering
\caption{\label{tbl:spec_params}
Fitted trend parameters ($\hat{m}_{j}^0$ and $\hat{\mu}_{j}$, errors are 1$\sigma$), values of the highest peak in power spectral density (PSD), corresponding periods ($P_{j}$) and false alarm probabilities (FAP) of these highest peaks. 
}
\begin{tabular}{cccccc}
	\hline
	\hline
		 & \multicolumn{5}{c}{Spectral analysis parameters}      \\
	Band & $\hat{m}_{j}^0$ & $\hat{\mu}_{j} \times 10^{-2}$   & $\max (\mathrm{PSD})$ & $P_{j}$ & FAP \\
		 &  & d$^{-1}$ &     & d &  per cent       \\
	\hline \noalign{\vskip 1.5ex}
	\multicolumn{6}{c}{OS 1} \\[1ex]
	$V$ &$\phantom{-}16.965 \pm 0.001$ & $\phantom{-}1.429 \pm 0.001$ & $0.114$ & $1.77$ & $10.35$\\
	$I$ &$\phantom{-}15.834 \pm 0.001$ & $\phantom{-}1.262 \pm 0.001$ & $0.072$ & $1.77$ & $26.89$\\
	$J$ &$\phantom{-}15.015 \pm 0.001$ & $\phantom{-}0.685 \pm 0.001$ & $0.145$ & $1.76$ & $71.21$\\
	$H$ &$\phantom{-}14.639 \pm 0.001$ & $\phantom{-}0.493 \pm 0.001$ & $0.059$ & $1.74$ & $57.46$\\[1.5ex]
	\multicolumn{6}{c}{OS 2} \\[1ex]
	$V$ &$\phantom{-}16.456 \pm 0.001$ & $-0.474 \pm 0.001$ & $0.054$ & $1.72$ & $76.21$\\
	$I$ &$\phantom{-}15.431 \pm 0.001$ & $-0.446 \pm 0.001$ & $0.048$ & $1.75$ & $79.91$\\
	$J$ &$\phantom{-}14.302 \pm 0.001$ & $-0.067 \pm 0.001$ & $0.188$ & $1.50$ & $13.97$\\
	$H$ &$\phantom{-}14.154 \pm 0.001$ & $-0.287 \pm 0.001$ & $0.053$ & $7.87$ & $63.30$\\[1.5ex]
	\multicolumn{6}{c}{OS 3} \\[1ex]
	$V$ &$\phantom{-}16.316 \pm 0.001$ & $\phantom{-}1.178 \pm 0.001$ & $0.245$ & $1.78$ & $\phantom{0}8.40$\\
	$I$ &$\phantom{-}15.291 \pm 0.001$ & $\phantom{-}1.006 \pm 0.001$ & $0.260$ & $1.78$ & $\phantom{0}0.93$\\
	$J$ &$\phantom{-}14.415 \pm 0.001$ & $\phantom{-}0.830 \pm 0.001$ & $0.252$ & $1.78$ & $\phantom{0}0.11$\\
	$H$ &$\phantom{-}14.074 \pm 0.001$ & $\phantom{-}0.719 \pm 0.001$ & $0.178$ & $1.78$ & $\phantom{0}0.23$\\[1.5ex]
	\multicolumn{6}{c}{OS 4} \\[1ex]
	$V$ &$\phantom{-}16.387 \pm 0.001$ & $\phantom{-}1.376 \pm 0.001$ & $0.268$ & $2.30$ & $\phantom{0}0.31$\\
	$I$ &$\phantom{-}15.392 \pm 0.001$ & $\phantom{-}1.159 \pm 0.001$ & $0.288$ & $2.30$ & $\phantom{0}0.20$\\
	$J$ &$\phantom{-}14.606 \pm 0.001$ & $\phantom{-}0.871 \pm 0.001$ & $0.119$ & $2.30$ & $35.63$\\
	$H$ &$\phantom{-}14.199 \pm 0.001$ & $\phantom{-}0.880 \pm 0.001$ & $0.170$ & $2.32$ & $44.58$\\[1.5ex]
	\multicolumn{6}{c}{OS 5} \\[1ex]
	$V$ &$\phantom{-}16.507 \pm 0.001$ & $-1.141 \pm 0.001$ & $0.295$ & $1.77$ & $\phantom{0}0.03$\\
	$I$ &$\phantom{-}15.469 \pm 0.001$ & $-1.113 \pm 0.001$ & $0.225$ & $1.77$ & $\phantom{0}0.11$\\
	$J$ &$\phantom{-}14.571 \pm 0.001$ & $-1.099 \pm 0.001$ & $0.200$ & $1.78$ & $\phantom{0}0.11$\\
	$H$ &$\phantom{-}14.182 \pm 0.001$ & $-0.871 \pm 0.001$ & $0.202$ & $1.77$ & $\phantom{0}0.06$\\[1.5ex]
	\multicolumn{6}{c}{OS 6} \\[1ex]
	$V$ &$\phantom{-}16.179 \pm 0.001$ & $-0.269 \pm 0.001$ & $0.214$ & $1.77$ & $\phantom{0}0.03$\\
	$I$ &$\phantom{-}15.144 \pm 0.001$ & $-0.262 \pm 0.001$ & $0.189$ & $1.77$ & $\phantom{0}0.05$\\
	$J$ &$\phantom{-}14.273 \pm 0.001$ & $-0.218 \pm 0.001$ & $0.132$ & $1.78$ & $\phantom{0}1.45$\\
	$H$ &$\phantom{-}13.963 \pm 0.001$ & $-0.245 \pm 0.001$ & $0.111$ & $1.78$ & $\phantom{0}0.47$\\
	\hline
	\end{tabular}
\end{table*}

Most of the light curves have only one data point per night, with rare exceptions of two subsequent observations in one night. 
For a strictly periodic observations (uniform time series) with a time step of 1 d, the highest (Nyquist) frequency is $0.5$~d$^{-1}$, corresponding to the smallest period of 2~d.
However, the analysed time series are not strictly uniform, because the time interval between the two subsequent observations is not precisely equal to 1 d. 
Instead, a typical interval ranges from 0.95 to 1.05 d, making light curves quasi-uniform time series with irregular gaps. 
As a result, we expect that a periodic modulation with $P \approx 1.77$~d will produce a peak in the PSD at frequency $\nu_1 = 0.565$~d$^{-1}$ and an alias at frequency $\nu_2 = 0.435$~d$^{-1}$, which corresponds to a period of $\approx 2.29$~d.
The relative amplitudes of these peaks depend on the profile of the window function, which in turn depends on the distribution of observation times within studied time interval and on the number of observations. 
In general, for a non-uniform time series, the amplitude of the true signal is expected to be larger than the amplitudes of its aliases.
We verify this property by studying the individual window function in each case.

\begin{figure*}
	\includegraphics[keepaspectratio, width = 0.48\linewidth]{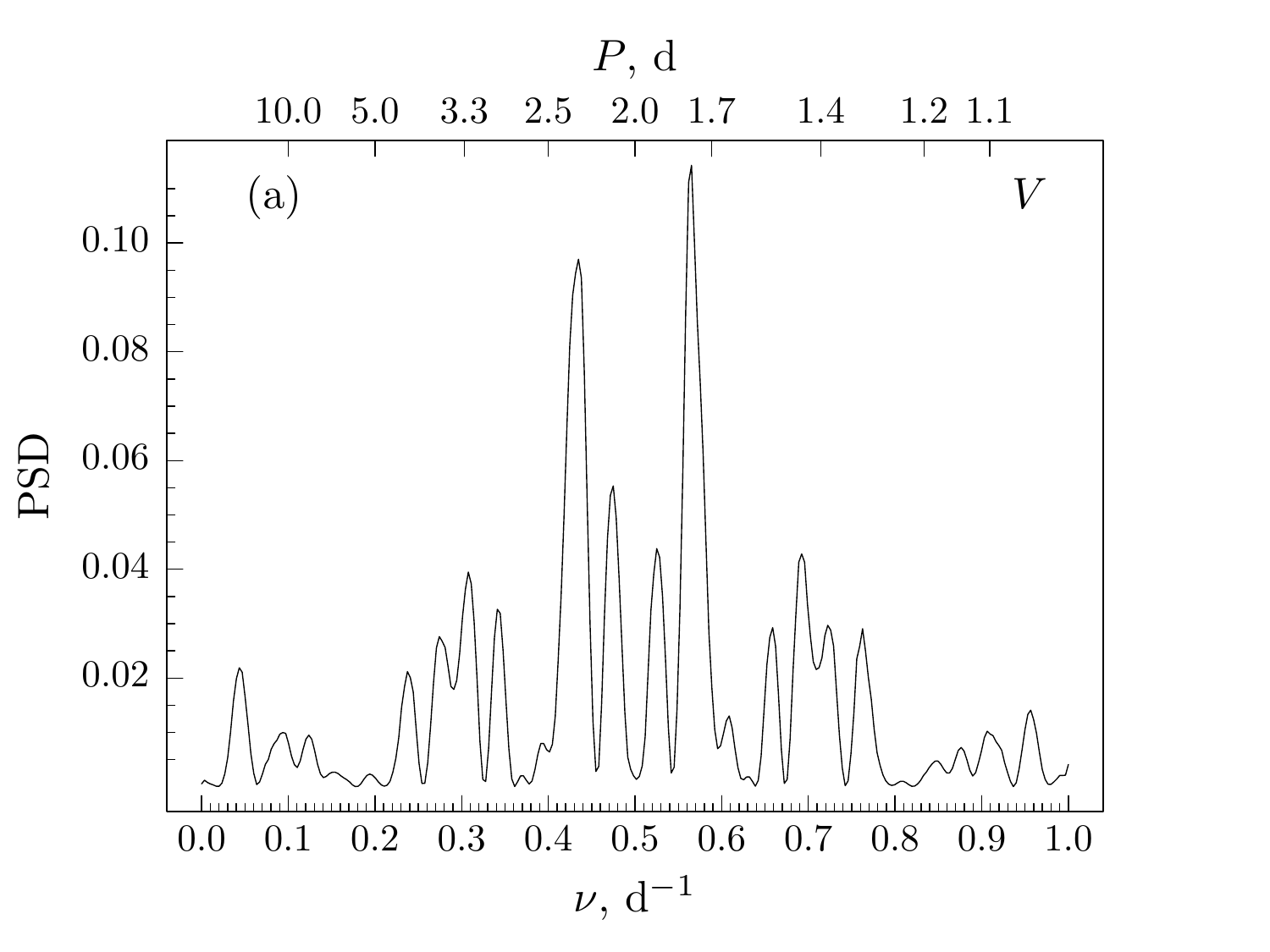}
	\hfill
	\includegraphics[keepaspectratio, width = 0.48\linewidth]{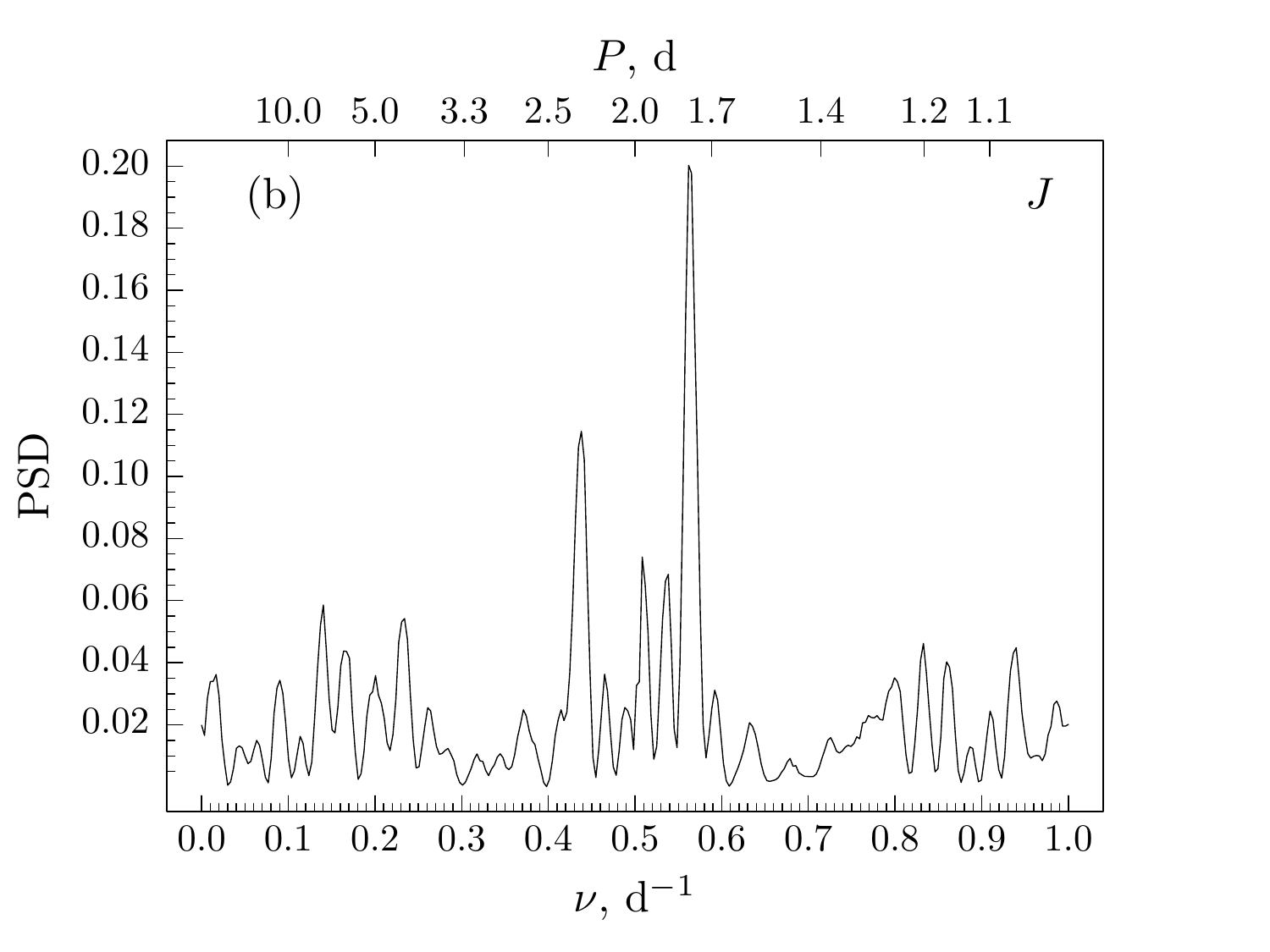}
\caption{\label{fig:prdgr}
Example power spectral densities calculated for (a) the \ti{V}\ band of \oi1 and (b) the \ti{J}\ band of \oi5. 
The values of the highest peaks, corresponding periods and false alarm probabilities are given in Table~\ref{tbl:spec_params}.
}
\end{figure*}

\begin{figure*}
		\includegraphics[keepaspectratio, width = 0.48\linewidth]{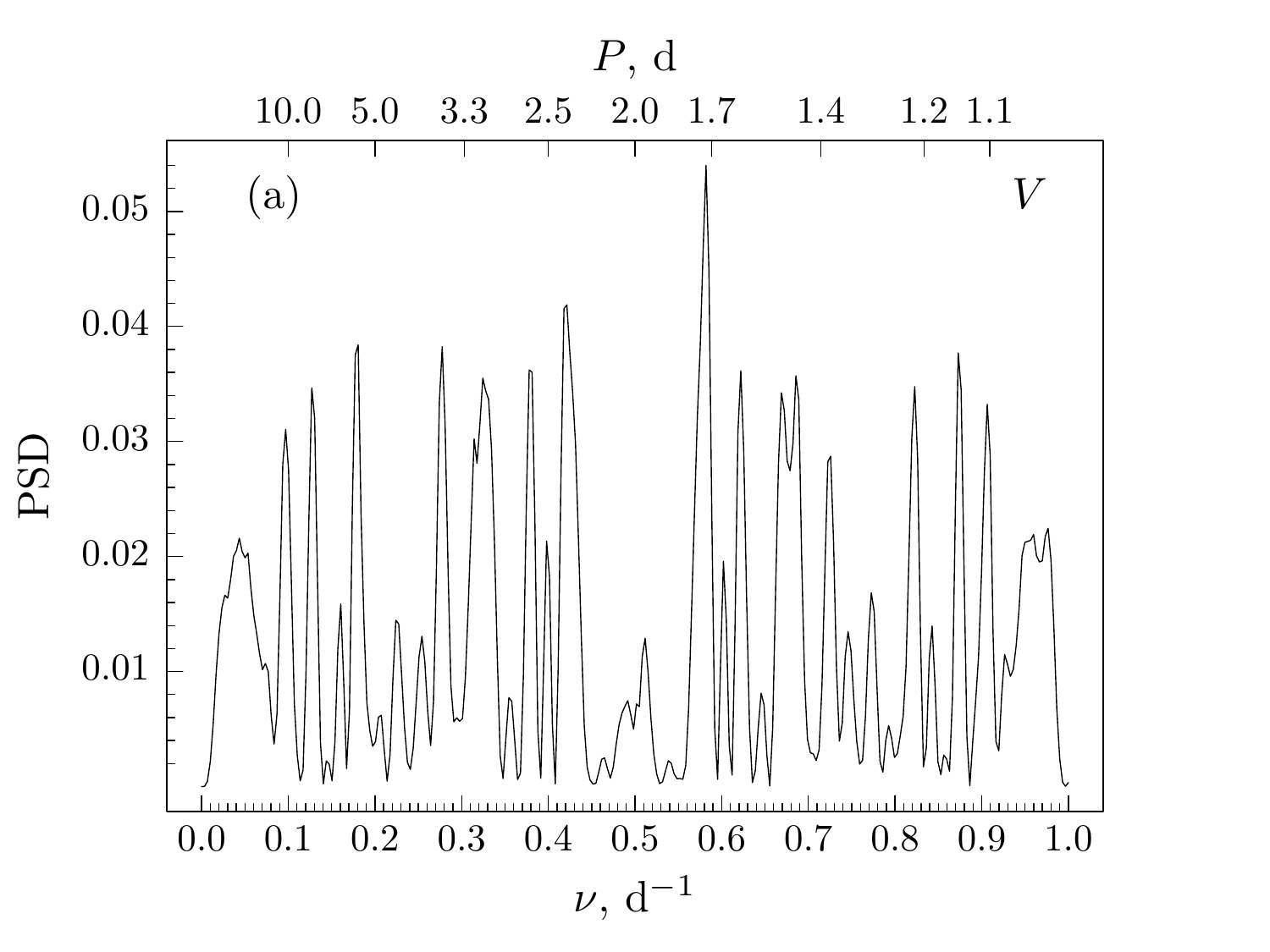}
	\hfill
		\includegraphics[keepaspectratio, width = 0.48\linewidth]{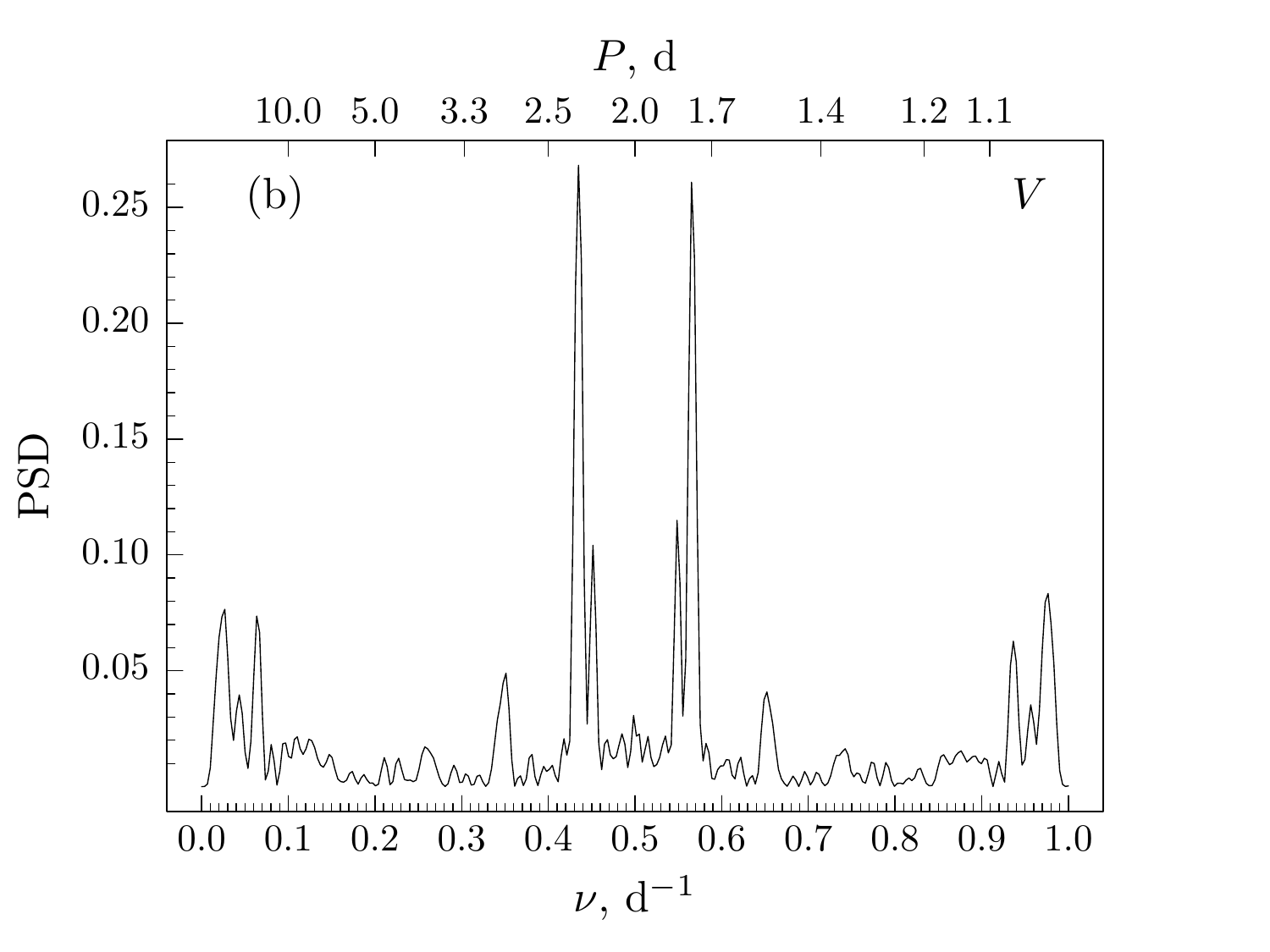}
\caption{\label{fig:prdgr_failed}
Examples of atypical profiles of power spectral densities (PSDs).
(a) PSD calculated for the \ti{V}\ band of \oi2. 
The highest peak could be spurious, which is indicated by the high false alarm probability of 76 per cent (see Table \ref{tbl:spec_params}). 
(b) PSD of the \ti{V}\ band of \oi4. 
The alias peak is higher than the real peak, unlike in other \oi{s}.
}	
\end{figure*}

In order to distinguish between spurious and real peaks, we estimate the false alarm probabilities (FAP) of the highest detected peak in each PSD. 
The FAP can be obtained from cumulative probability distribution (CDF) of maxima of Lomb-Scargle PSDs, calculated for the time series with no periodic component. 
A number of analytical formulae describing CDFs were derived for uniform series \citep[see e.g.][]{Scargle1982}, but no simple analytical solution was found for a general non-uniform series, and the CDFs are usually constructed using numerical simulations. 
In order to estimate the FAP of signals detected in the PSDs of \GX, we adopt the numerical scheme described in \citet{Frescura2008}. 
This method involves simulating a large number of test (random) time series with the same window function, and calculating the maxima of the corresponding PSDs. 
The obtained values are then used to construct an approximation of the CDF, which in turn provides an estimate of FAP for the maximum peak in the PSD of the observed time series. 
After some investigation, we decided to limit the number of test time series to 10$^4$ for each band of each observation interval. 
The resulting FAPs, amplitudes of maximum peak in the PSDs and corresponding periods are presented in Table~\ref{tbl:spec_params}.
We note that the FAPs were calculated for each light curve independently, not taking into account the probability of detecting the peak at particular frequency in each light curve.

Fig.~\ref{fig:prdgr} shows typical PSDs obtained for the \ti{V}\ band of \oi1 (panel a) and the \ti{J}\ band of \oi5 (panel b). 
Two clearly distinguishable peaks can be seen in each panel. 
The highest peak in these two cases ($\nu \approx 0.56$~d$^{-1}$, or $P \approx 1.77$~d) corresponds to the actual short-term variability found in the observed light curves of \GX, while the other one ($\nu \approx 0.44$~d$^{-1}$, $P \approx 2.28$~d) is caused by aliasing. 
The relative amplitudes of the signal peak and its alias differ between \oi{s} and bands, depending on the data spread and on the number of data points. 
The FAPs of the highest peaks of the PSDs in Fig.~\ref{fig:prdgr} are 10.4 and 0.1 per cent, respectively, indicating that in the \ti{V}\ band of \oi1 the periodic component is present, but that there is a high probability of this peak being caused by a coincidence, while for the \ti{J}\ band of the \oi5, it is highly unlikely that the observed peak is spurious. 
PSDs of other \oi{s} resemble those shown in Fig. \ref{fig:prdgr}, with the one exception of \oi2 (see Fig.\ref{fig:prdgr_failed}a). 
Both the Bayesian fitting procedure and the Lomb-Scargle method failed to detect any significant periodic component, with period close to the orbital period, in all ONIR bands of \oi2. 

An atypical PSD shape was obtained for \oi4 (see Fig.~\ref{fig:prdgr_failed}b). 
The highest peak is found at the alias frequency ($\nu \approx 0.44$~d$^{-1}$), although the peak at the true frequency is only slightly lower. 
The FAPs of these peaks in the \ti{V}\ and \ti{I}\ bands (see Table \ref{tbl:spec_params}) are low, indicating that the observed light curves indeed contain periodic components.
However, FAP calculated for \ti{J}\ and \ti{H}\ bands are relatively high, which in turn means that the periodic component in these two bands cannot be reliably identified. 
Although the highest peak is found at the alias frequency, it can still be used to estimate the properties of the periodic component, at least in the \ti{V}\ and \ti{I}\ bands.

\section{Discussion}\label{sec:Disc}
\label{sec:ConDis}

\begin{figure}
	\includegraphics[keepaspectratio, width=1\linewidth]{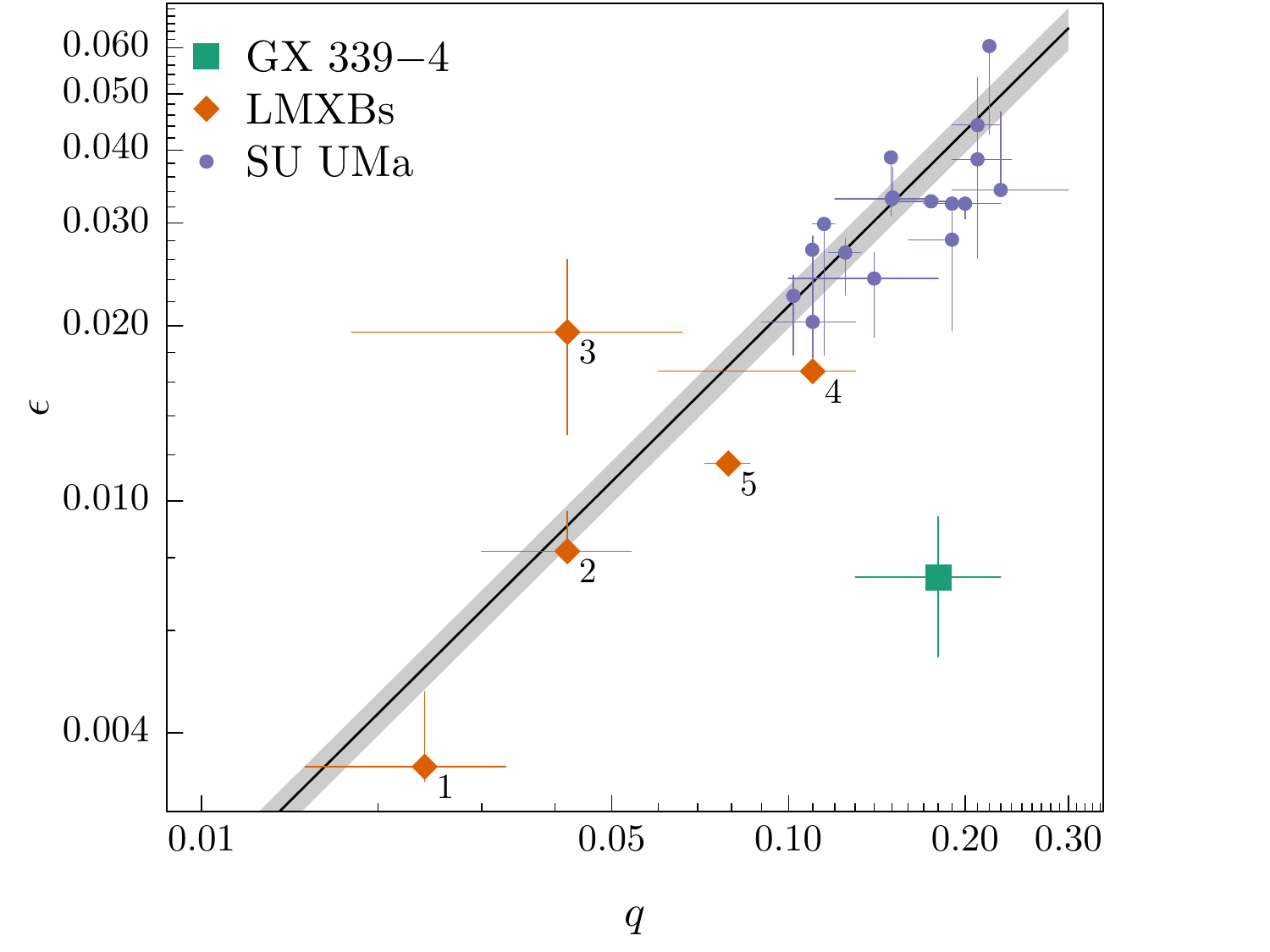}
	\caption{\label{fig:shump} Period excess of superhumps as a function of mass ratio. Data are adopted from \citet{Smith2007} (see their table 5 and references therein). Binary mass ratios of LMXBs are taken from \citet{Casares2014} (see their table 1 and references therein) and from \citet{Wu2015} for N Mus 1993. Errors are 1$\sigma$.
	The labeled LMXBs are: (1) KV UMa (XTE J1118$+$480), (2) Qz Vul (GS 2000$+$2), (3) V1482 Aqr (GRS 1915$+$105), (4) V518 Per (GRO J0422$+$32), (5) GU Mus (N Mus 1991).}
\end{figure}

\begin{table*}
\centering
\caption{\label{tbl:qscnt}
List of start and end dates, the number of observations, and the average magnitudes with standard deviations for the analysed quiescent states.
}	
\begin{tabular}{cccccccccc}
\hline
\hline
 Start date      & End date    & \multicolumn{4}{c}{N of observations} & \multicolumn{4}{c}{ Average magnitude }  \\
 MJD$\ -\ 50000$ & MJD$\ -\ 50000$ & $V$ & $I$ & $J$ & $H$  & $V$ & $I$ & $J$ & $H$ \\
\hline
 $2856.60$ & $2943.50$ & 31 & 27 & 33 & 41 & 19.65 $\pm$ 0.06 & 18.31 $\pm$ 0.03& 17.24 $\pm$ 0.12 & 16.67 $\pm$ 0.07 \\
 $3560.70$ & $3660.50$ & 20 & 22 & 27 & 33 & 18.94 $\pm$ 0.12 & 17.63 $\pm$ 0.01 &  16.54 $\pm$ 0.02 & 15.96 $\pm$ 0.02 \\
 $3968.59$ & $4019.51$ & 13 & 17 & 9  & 14 & 19.44 $\pm$ 0.04 & 17.85 $\pm$ 0.07 & 16.71 $\pm$ 0.04 & 16.17 $\pm$ 0.15 \\
 $5700.62$ & $5735.62$ & 18 & 20 & 19 & 20 & 19.36 $\pm$ 0.20 & 18.03 $\pm$ 0.21 & 16.82 $\pm$ 0.17 & 16.31 $\pm$ 0.28 \\
\hline
\end{tabular}
\end{table*}

In order to investigate the origin of the reported variability, we apply Lomb-Scargle spectral analysis to the data in the quiescent state.
The observed flux of \GX\ during its faintest flux periods was contaminated by the flux from the nearby field stars \citep{Buxton2012}. These observations are included in the publicly available SMARTS data, but we did not use these data in our analysis.
The dates for the selected intervals are listed in Table~\ref{tbl:qscnt}.
We found that the false alarm probabilities of the highest PSD peaks range from 30 to 90 per cent, and the corresponding frequencies of the peaks appear to be random.

The contribution of the secondary to the total NIR flux is estimated to reach 50 per cent in quiescence \citep{Heida2017}, and hence the donor star contributes approximately 5 per cent during soft states, when the total NIR luminosity is about an order of magnitude larger. 
However, the X-rays originating from the vicinity of the compact object can irradiate the surface of the donor star and increase its contribution to the total flux in the soft state. 
We estimate this contribution by considering the X-ray luminosity to be 10 per cent of the Eddington limit for a 5~M$_{\sun}$ black hole \citep{Hynes2003, Heida2017}, and ba assuming the X-ray emission pattern to follow the Lambert's law (proportional to the cosine of the inclination angle between the direction of outgoing emission and the disc axis; see, however, a more precise approximation in \citealt{Suleimanov2008}). 
The disc covers a fraction of the secondary surface, its opening angle is assumed to be 12$\degr$ \citep{deJong1996} and binary mass ratio is 0.18 \citep{Heida2017}. 
We take the distance to the source equal to 10~kpc \citep{Hynes2003, Heida2017} and the interstellar reddening to be A$_V=3.7$ \citep{Zdziarski1998, Buxton2012}.
We obtain the flux from the irradiated surface of the donor to be of the order of 0.01~mJy in the \ti{V} band, much smaller than the observed flux of $\sim 0.5$~mJy from the source in the soft state. 
We conclude that the soft state emission is dominated by the accretion disc, which is also responsible for the observed variability.

The small (about 1 per cent) excess above the orbital period and the absence of oscillations during quiescence suggest that the observed variability is caused by the superhumps.
Superhumps are optical periodic modulations that were originally observed in SU UMa dwarf novae \citep{Vogt1974, Warner1975}. 
Superhumps usually accompany superoutbursts and are never observed during normal outbursts or quiescent states of dwarf novae \citep{Osaki1996}.
These modulations are believed to be caused by the slow precession of an eccentric accretion disc, which is deformed owing to the presence of the 3:1 resonance within it \citep{Whitehurst1991}. 
The prograde precession of the disc leads to the observed superhump period, $P_\mrm{sh}$, being slightly larger (by a few per cent) than the orbital period, while a more rare retrograde precession causes the observed superhump period to be smaller \citep{Wood2011}. 
The actual period of the disc precession is usually much longer than the orbital period, and can be expressed as $P_\mrm{prec} = {P_\mrm{sh}}/{\epsilon}$, where $\epsilon = (P_\mrm{sh} - P_\mrm{orb}) / P_\mrm{orb}$ is the superhump period excess \citep{Haswell2001}. 
The period excess can in turn be expressed as \citep{Osaki1985,Mineshige1992}
\begin{equation}\label{eq:epsilon}
\epsilon = \frac{1}{4}\frac{q}{\sqrt{1+q}}\eta^\frac{3}{2},
\end{equation}
where $q = M_2/M_1$ is the mass ratio and $\eta = R_\mrm{d}/R_\mrm{crit}$ is the ratio of the disc radius to the critical radius, beyond which the disc becomes unstable \citep{Hirose1990}.

Superhumps have also been detected in black hole transient LMXBs \citep{Kato1995, Donoghue1996, Uemura2000, Zurita2002,Zurita2008}.
One of the most plausible explanations of LMXB superhumps is that changes in both the are of the disk visible to the observer and the modulation of the fraction of intercepted X-ray emission over the superhump cycle contribute to the observed optical modulations \citep{Haswell2001}.
The 3:1 resonance condition restricts the mass ratio $q \le 0.25$ \citep{Whitehurst1991} of the systems demonstrating superhump modulations. 
The condition is typically fulfilled in LMXBs with massive primaries \citep{Casares2014}, and is also true for \GX, for which $q=0.18\pm0.05$ was recently measured \citep{Heida2017}. 
We estimate a superhump period excess for \GX\ of $\epsilon = 0.007\pm0.002$, which is smaller than the typical values obtained for cataclysmic variables with the same mass ratio (see Fig.~\ref{fig:shump} and \citealt{Smith2007}). 
For this $\epsilon$, the disc precession period of \GX\ is $P_{\rm prec}=240$~d.

In Fig.~\ref{fig:shump} we show that the analytical $\epsilon-q$ relationship (Eq.~\ref{eq:epsilon}, solid black line) aligns well with the observed dwarf novae if we put $\eta \approx 0.9$,  but LMXBs tend to have smaller superhump period excesses. 
Smaller values of $\eta$ could potentially account for this discrepancy. 
In order to explain the superhump excess observed in \GX\, $\eta = 0.1$ is required.
The physical reasons for the accretion disc in \GX\ being substantially smaller are not clear.

\section{Conclusions}

We investigated the variability of the long-term $V$, $I$, $J$ and $H$ light curves at periods close to the orbital period.
We chose six intervals of observations away from the flares, which coincide with the soft state of the source.
We used two different methods, namely Bayesian inference and Lomb-Scargle spectral analysis, and found prominent oscillations in five intervals (\oi1, 3, 4, 5 and 6) at $P\approx1.772$~d, while the spread in the data points of \oi2 does not allow reliable estimation of the periodic oscillations.
We additionally considered four intervals corresponding to the quiescent state and found that none of these light curves demonstrates significant oscillations.
We conclude that the observed oscillations appear during the soft states and probably originate from the accretion disc.

The calculated periods indicate, despite the long time gaps between subsequent soft states, that the detected period is fairly stable (Fig.~\ref{fig:bar_p_ph}, left panel).  The spread of periods in different filters within one \oi\ is typically much smaller than the spread of periods for different \oi{s}.

Despite the spread of the periods in the soft-state data, the determined periods (those with small error bars, e.g. in the $V$ and $I$ filters) are systematically larger than the values obtained for the orbital period of the system, 1.7557 $\pm$ 0.0004~d \citep{Hynes2003} and 1.7587 $\pm$ 0.0005~d \citep{Heida2017}.
We obtained the average period of $P=1.772$~d (see Fig.~\ref{fig:bar_p_ph}, solid black line) and the standard deviation of the period distribution is $0.003$~d for the five observational sets with the prominent periodic component.
The difference between the orbital period and the measured periods significantly exceeds typical measurement errors in 17 out of 20 cases.
Such high accuracy in period estimation was only possible thanks to the exceptionally long observations of the system, despite the small number of points in each period.
We compared the superhump excess period to other systems, LMXBs and SU UMa dwarf novae, and found that the excess in \GX\ is substantially below than that expected for the binary with the same mass ratio.
The physical reasons for that are, however, not clear.

\section*{Acknowledgements}

This paper has made use of SMARTS optical/near-infrared light curves that are available at \url{www.astro.yale.edu/smarts/xrb/home.php}.
IK acknowledges Nordita Visiting PhD Fellow program.
AV acknowledges support from the Academy of Finland grant 309308 and the Ministry of Education and Science of the Russian Federation grant 14.W03.31.0021.
We thank Juri Poutanen, Vitaly Neustroev, Valery Suleimanov, Emrah Kalemci, Jorge Casares and Tolga Din\c{c}er for helpful discussions and the referee for the comments which greatly improved the manuscript.




\bibliographystyle{mnras}
\bibliography{references}






\bsp	
\label{lastpage}
\end{document}